\def \kpc {\rm kpc}
\def \kms {\rm km~s^{-1}}
\documentstyle [referee,psfig]{l-aa}

\begin{document}

\thesaurus{4(8.11.1; 10.06.2; 10.11.1; 10.19.1; 10.19.3)}

\title{Stellar kinematics in the solar neighbourhood \\
and the disc scale lengths of the Galaxy}

\author{O.~Bienaym\'e and N.~S\'echaud }

\offprints{O.Bienaym\'e}

\institute{
Observatoire de Strasbourg, CNRS URA 1280, 11 rue de l'Universit\'e,
F-67000 Strasbourg, France
}

\date{Received 7 May 1996, accepted January 1997}

\maketitle

\markboth {O.~Bienaym\'e \& N.~S\'echaud:
 Stellar kinematics in the solar neighbourhood } {}

\begin{abstract}

A general dynamically consistent 2D flat distribution function is
built to model the kinematics of neighbouring stars. Application leads
to the measurement of a short galactic scale length $R_\rho$ between
1.7 and $2.9\,\kpc$ and a
locally decreasing rotation curve.  This is in agreement with recent
determinations based on kinematics and counts of distant stars.  These
results rule out the classical assumption that $2R_\rho=R_\sigma$ or
that $\sigma_z(R)/\sigma_R(R)$ is constant when the vertical scale
height $h_z(R)$ is assumed to be constant.  We explain why the
measured squared axis ratio of the velocity dispersions
$\sigma_v^2/\sigma_u^2$ of disc stars is less than $1/2$.  This ratio
has been claimed to be important evidence for the non-axisymmetry of
the galactic disc.  We show that this is not the case and that it may be
simply explained with a realistic axisymmetric disc model if the
circular velocity is locally declining or if there is a mismatch
between the photometric and kinematic scale lengths.


\end{abstract}

\section{Introduction}
Dynamical self-consistent constraints deduced from the Boltzmann equation
are necessary to analyse and interpret kinematic data in the Galaxy.
It has been shown that the standard epicycle theory approach, a first
order theory to estimate galactic characteristics, is subject to
systematic errors (see Kuijken \& Tremaine 1991, who put forward
higher order developments).
In some recent work, distribution function solutions of the Boltzmann equation
(proposed by Shu 1969) and/or third moments in Jeans equations (Cuddeford
\& Amendt 1991) have been used to obtain a more consistent description of
our Galaxy (Cuddeford \& Binney 1994, Kuijken \& Tremaine 1991, Evans \&
Collett 1993, Fux \& Martinet 1994).
With different approaches, they attempt to determine and to make use of exact
constraints between global parameters describing our Galaxy, namely the slope
of the circular velocity curve, the density and kinematic scale lengths etc...

In this paper, we give a short review of observational constraints and
the different values taken by these authors.  Secondly we build
a new model of Shu-type distribution functions
where scale lengths and the shape of the velocity
curve are free parameters.  Thirdly we compare and adjust this model
to velocity distributions of stars in the solar neighbourhood taken from the
Gliese catalogue and other catalogues.  Implicit constraints are related to the
asymmetric drift relation and the Lindblad equation that gives the
ratio of radial and tangential velocity dispersions.  We conclude that
extended Shu distribution functions
recover most of the fundamental kinematic properties
of the galactic disc.  We show that it is possible to build a
consistent model explaining local stellar kinematics and
to determine local structural galactic parameters.

\section{Observational constraints}
We use local kinematic data from Gliese \& Jahreiss (1991) and other
catalogues to constrain galactic structure. Stars near the sun with large
relative velocity provide information on non-local galactic structure like
the kinematic and density gradients or the slope of the rotation curve.
The local star velocity distribution is a section of phase space and this
observed distribution depends on non-local quantities like the potential,
or global density distribution of stars (other examples are given by Dehnen
and Binney 1995, who describe ways to map the Galaxy's gravitational potential
and distribution of matter).
We show that the analysis of local kinematics is able to recover the
potential gradient and stellar scale lengths.
In previous analyses, assumptions have been either a flat velocity curve
(Evans \& Collett 1993) or a relation between density and kinematic gradients
(Cuddeford \& Binney 1994).
These assumptions allow these authors to obtain tractable distribution
functions. However they found that these distribution functions
are not compatible with existing observational evidence.
We discuss observational evidence on galactic structure, and emphasize the
existing systematic differences between authors.

\subsection{The circular velocity curve}
Determinations of the galactic circular velocity curve are based on radial
velocity observations of objects belonging to the disk population. Due to
their small velocity dispersion, they follow the galactic rotation curve
closely. However exact determination of the velocity curve depends critically
on the adopted galactic radius $R_0$ and circular velocity $\Theta_0$ at the
solar position. Assuming $R_0=7.9\,\kpc$ and $\Theta_0=185\,\kms$, Rohlfs and
Kreitschmann (1988) found that the velocity curve ``is slowly declining from
$\Theta=200\,\kms$ at  $R=4.5\,\kpc$ to $\Theta=173\,\kms$ at $R=11\,\kpc$''
(however taking $\Theta_0=220\,\kms$, they obtained a flat rotation curve).
This result is also corroborated by Fich et al. (1989), Pont et al. (1994)
and by Dambis et al. (1995).

Since most recent determinations favor values around $R_0=7.9\,\kpc$
(Reid 1993) or $R_0=8.1\,\kpc$ (Pont et al. 1994) and
$\Theta_0=185\,\kms$, much smaller than the recommended values of
$R_0=8.5\,\kpc$ and $\Theta_0=220\,\kms$ by the IAU general assembly in
1985 (Kerr \& Lynden-Bell 1986), we consider it very likely that
the circular velocity curve is locally declining. We estimate $\alpha
\simeq -0.1\,\mbox{to}\,-0.3$ for $v_c(R)=R^{\alpha}$ at the solar galactic
radius of $R_0=7.5\,\kpc$ and $\Theta_0=185\,\kms$.

Considering the Lindblad equation
$\sigma_\phi^2/\sigma_R^2\simeq{1\over2}(1+{d\ln~v(R) \over d \ln R})$, we
conclude that there is no a priori conflict between the slope of the rotation
curve and the observed ratio of velocity dispersion
$\sigma_\phi^2/\sigma_R^2$ smaller than ${1\over2}$ (Kerr \& Lynden-Bell 1986).

\subsection{The disc density scale length}
Most disc density scale length determinations range between 2.5 and
$5\,\kpc$ (see Kent et al. 1991 and Robin et al. 1992a), but part of
these measurements are model dependent and require
assumptions. Assuming $\sigma_R(R)=\exp(-R/2h)$, Lewis \& Freeman
(1989) gave $h=4.4\,\kpc$ from velocity dispersions of distant K
giants. van der Kruit (1986) obtained a larger value of $5.5\,\kpc$
from the $Pioneer~10$ background experiment. But his determination
gives access only to the radial to vertical scale ratio, and he
assumed a $h_z=325~pc$ vertical scale height for the old
disc. Adopting the recent determination of 250 pc for $h_z$ (Kuijken
\& Gilmore 1989, Haywood et al. 1996), the radial scale length deduced
from $Pioneer~10$ data would be closer to $4.2\,\kpc$. Recent direct
determinations range between $2.5\,\kpc$ (star counts: Robin et
al. 1992ab, Ojha et al. 1996, COBE map: Durand et al. 1996) and about
$3.5\,\kpc$ (Kent et al.  1991). We note that the kinematic
determination of the radial scale length obtained by Fux \& Martinet
(1994) is $h=2.5\,\kpc$, assuming a radially constant scale height
$h_z$, and $h=3.1\,\kpc$ with a positive local $h_z$ gradient of
$30\,pc/\kpc$.  Their analysis is based on the asymmetric drift
equation and on moments of the Boltzmann equation.

\subsection{The disc kinematic scale length}

We know only few direct determinations for the kinematic scale length of
stellar discs in the Milky Way. Neese \& Yoss (1988) measured the gradient of
radial velocity dispersion from 364 stars mainly towards the galactic
anticentre and found
$\partial \sigma_R/\partial R=-3.8\pm0.6\,\kms\,\kpc^{-1}$ or
$\partial \ln\sigma_R^2/\partial R= -(3.7\,\kpc)^{-1}$.
Lewis \& Freeman (1989) obtained $4.4\,\kpc$ for the $\sigma_R^2$ scale length
(or $8.8\,\kpc$ for $\sigma_R$). These results are based on 600 distant giants
towards the galactic centre and anticentre. As remarked by Evans \& Collett (1993),
the Lewis \& Freeman (1989) fit is probably not good at the solar position and a
different value for the velocity dispersion at the solar galactic radius based on K
giants (Delhaye 1965) should be used. With this value Evans \& Collett (1993)
tried a non-exponential function to model the radial dependence of velocity
dispersions and obtained a better fit. Using this function, we obtain for the
local kinematic scale length $\partial \ln\sigma_R^2/\partial
R= (-4.8\,\kpc)^{-1}=-0.21\,\kpc^{-1}$.
This illustrates the precision of the determination. From star counts
and proper motions Ojha et al. (1996) measured velocity dispersion and
gradients and obtained similar kinematic scale lengths.

It is usually accepted that galactic discs have constant scale heights
and exponentially decreasing velocity dispersions.  Measured velocity
dispersions in galactic discs led Bottema (1993) to conclude that the
velocity dispersion decreases exponentially with radius like
$\sigma_R^2 \sim e^{-R/R_\rho}$ if the vertical to radial velocity dispersion
ratio is constant with radius.  This last current assumption (or the
plane-parallel assumption) must be ruled out if, for example, the
observations in the Milky Way show that $R_\rho \neq R_{\sigma^2}$.

For a given disc galaxy, there is not a strong chromatic dependence of
the scale length (Elmegreen \& Elmegreen 1984, Giovanardi \& Hunt
1988). The bulk of 86 spiral galaxies (Fig. 9 by de Jong, 1996) shows
$R_B/R_K \sim1.2$. This is a strong indication that the
stellar populations are distributed in discs with similar scale
lengths.  For our model fitting in the next section, we will assume no
chromatic dependence for our Galaxy.  We will assume the same for
kinematic scale lengths though there are no existing observations that
confirm this.

\subsection{Disc axisymmetry}

Evidence from a variety of sources shows the existence of a bar in the
central few kpcs of the galactic center, with rough agreement on the
orientation of the bar. However it is much more difficult to find
evidence from star counts in the galactic plane for ellipticity of the
stellar disc at the solar galactic radius, particularly if we are on
one of the symmetry axes.

Signatures from kinematics give more reliable constraints (Kuijken \&
Tremaine 1994): 1) Oort's constants C and K and the LSR radial velocity
show no evidence for local non-axisymmetry, 2) while vertex deviation for
low velocity dispersion and young stars may show such evidence
this last effect could be caused by very local non-stationary
effects related to the presence of spiral arms, 3) for stars in the solar
neighbourhood of high velocity dispersion, the velocity ellipsoid points
towards the galactic centre ($l=5.5^\circ\pm4.2$), with no indication of local
non-axisymmetry, 4) the axis ratio $\sigma_v / \sigma_u $ is
observed to be about $\sim 0.5$ and can be a signature of ellipticity
but as we show in detail in this paper, it may also be explained
if the rotation curve is locally decreasing or if the kinematic and
density scale lengths are very different.

Stronger arguments in favour of the effect of a bar at $R=R_0$ come
from the Blitz \& Spergel (1991) analysis of HI data from which they
deduced that the LSR has a radial motion of $14\,\kms$. However the
effect of the warp beyond $R=12\,\kpc$ (Burton \& te Lintel Hekkert
1986) may modify their analysis and the LSR radial motion is not
confirmed by other observational measurements:
$v_r(LSR)=-1\pm9\,\kms$ (Table 3 in Kuijken \& Tremaine 1994). As a
conservative hypothesis we will still assume that the galactic disc is
axisymmetric and we will show that local kinematics may be
explained in the frame of this hypothesis.

\subsection{The vertical structure}
A plane-parallel potential is developed in the next section. Such
potentials do not permit the consideration of the correlations between vertical
and radial motions that must exist. From the probable mass galactic
distribution, these correlations have certainly a better
representation with a spherical potential than with a plane-parallel
one (this is not true for the density distribution). Such a spherical
potential will be briefly developed in the next section and general
solutions fitting observational constraints will be given in Section 4.
Intermediate galactic potentials between plane parallel and spherical
do not have 3 known integrals of motion (with the exception of St\"ackel
potentials) and cannot be used to find simple analytic distribution
functions. For this reason we restrict the present analysis to
St\"ackel potentials to include the effect of the vertical structure.
In these cases, the distribution functions are deduced following Statler (1989)
and are extensions of the Shu and Schwarzschild distribution functions.

Vertical structure introduces a bias on the Lindblad and asymmetric
drift equations which is estimated by Cuddeford \& Binney (1994). The
complete analysis done by Fux and Martinet (1994) allows them to
measure locally the third order derivative term of the potential
related to the flatness of the potential. This derivative is related
to the radial gradient of the vertical scale height $h_z$.

\section{Galactic disc distribution function}
Shu (1969) described a phenomenological distribution function (DF) that
has the great advantage of relative simplicity and that has been used
frequently.  This distribution function is a two dimensional
representation of a flat and rapidly rotating stellar disc in an
axisymmetric potential (rotational symmetry). The DF is written
explicitly in terms of two integrals of motion (energy and angular
momentum) and is a steady-state solution of the Boltzmann equation
if the potential is time-independent. For small velocity dispersions,
Shu's DF closely approximates the Schwarzschild velocity ellipsoid
DF. Two functions are free: the input surface density distribution
$\widetilde{\Sigma}(R)$ and the input radial velocity dispersion
$\widetilde{\sigma}(R)$ (Shu 1969, Tamisier 1991). The main advantage
of the Shu DF is that the associated exact surface density distribution
$\Sigma(R)$
and radial velocity dispersion $\sigma_R(R)$ (DF moments of 0 and $2^{nd}$
order) will in general remain close to the input functions
$\widetilde{\Sigma}(R)$ and $\widetilde{\sigma}(R)$, even with quite
large velocity dispersions. We define the Shu DF as
\noindent
\begin{eqnarray}
f\left(E,L_z\right)= {2\Omega\left(R_c\right)\over \kappa\left(R_c\right)}
{\widetilde{\Sigma}\left(R_c\right)\over 2\pi \widetilde{\sigma}^2\left(R_c\right)}
\exp[-{ \left(E-E_{circ}\left(R_c\right)\right) \over \widetilde{\sigma}^2\left(R_c\right)} ]
\mbox{~~~ if~~~ } L_z > 0
\end{eqnarray}
\begin{equation}
f(E,L_z)= 0~~ \mbox{ if } ~~ L_z \leq 0,
\end{equation}
where $R_c=R_c(L_z)$ is the radius of the circular orbit with angular momentum
$L_z$, ($R_c(L_z)$ is bijective for potentials considered in the next
sections), $E_{circ}(R_c)$ is the energy of a circular orbiting star at radius
$R_c$ with angular momentum $L_z$.
$\Omega$ is the angular velocity and $\kappa$ is the epicyclic frequency
$2\Omega[1+{1\over2}d~\ln\Omega/d~\ln~R]^{1/2}$.
The exponential part of the DF (Eq.1) represents a ring of stars with angular
momentum $L_z$ rotating near the radius $R_c$ and with radial velocity
dispersion $\widetilde{\sigma}$. The other terms are normalizations which
permit a distribution of elementary rings with different $L_z$ (and
$R_c$) and with a total mean density close to $\widetilde{\Sigma}(R)$.
Equation 2 is given to ensure that the total angular momentum of the disc is
non-zero. In this model the disc has a maximum rotation compared to
any other distribution function with the same resulting density and
no retrograde orbits are allowed. Here these restrictions have no consequence
since we analyse the local kinematics where only one retrograde orbit is found
among one thousand stars.

Since we want to build a model whose resulting density is
$\Sigma(R) \sim \Sigma_0 \exp(-R/R_\rho)$, we put
\begin{equation}
\widetilde{\Sigma}(R_c(L_z)) = \Sigma_0 \exp[-R_c(L_z)/R_\rho]
\end{equation}
and in order to have a nearly exponential $\sigma_R(R)$ distribution, we put
\begin{equation}
\widetilde{\sigma}(R_c(L_z)) = \sigma_0 \exp[-R_c(L_z)/R_\sigma].
\end{equation}

Thus the DF depends only on the integrals of motion and on the following free
quantities: the velocity curve $v_c(R)$ and the two constant scale lengths
$R_\rho$ and $R_\sigma$.

\subsection{Flat rotation curve model}
We describe a model with a flat rotation curve $v_c(R)=1$ and a
potential given by $\Phi(R)=\ln(R)$. Then we have
$R_c(L_z)=L_z/v_c(R_c)=L_z$. Pertinent parameters are the amplitude
$\sigma_0$ and the scale length ratio $R_\sigma/R_\rho$.
We define models with quasi-exponential density and dispersion
distributions:
\begin{equation}
\widetilde{\Sigma}(R_c(L_z)) = \Sigma_0 \exp[-L_z/R_\rho]
\end{equation}
and
\begin{equation}
\widetilde{\sigma}(R_c(L_z)) = \sigma_0 \exp[-L_z/R_\sigma].
\end{equation}

For different values of $\sigma_0$, Figs. 1a-2a show that the density
$\Sigma(R)$ and the radial velocity dispersion $\sigma_R(R)$
remain close to an exponential (with scale lengths $R_\rho$ and $R_\sigma$
respectively) for models with the smallest velocity
dispersions.  For the largest dispersion ($\sigma_0=.64$), the
effective density scale length is larger than the $R_\rho$ parameter.
The mean streaming velocity $\overline{v}_\phi$ (Figs. 1b-2b) of the
modelled stellar populations departs from the circular velocity curve
($v_c(R)=1$). This is due to asymmetric drift which is
directly related to the amplitude of the radial velocity
dispersion. For the models with large $\sigma_0$, the ratio
$\sigma_\phi/\sigma_R$ (Figs. 1c-2c) departs from the limiting value
$\sqrt{2}/2$ for zero velocity dispersion.

These results can be directly compared to that of Evans \& Collett (1993) and
Cuddeford \& Binney (1992): the velocity dispersion ratio can be much smaller
than $\sqrt{2}/2$, as observed in the solar neighbourhood (Table 8 of Kerr \&
Lynden-Bell 1986) in at least three situations: $R_0$ is smaller than
$R_\rho$, $R_\sigma$ is much larger than $R_\rho$ or the rotation curve is
decreasing with radius.

For small velocity dispersions, the ratio $\sigma_\phi/\sigma_R$ depends only
on the shape of the velocity curve. This is not true for large dispersions,
where dependence on the scale lengths is significant.

Shu-type models exist with the  Lewis \& Freeman (1989) assumption that
$2R_\rho=R_\sigma$ (Fig. 1), (see for example Binney 1987, Kuijken \&
Tremaine 1991 and Cuddeford \& Binney 1992). Such models imply  that
$R_\rho>5\,\kpc$ (i.e. $R_0/R_\rho<1.6$) to explain measured ratios
$\sigma_\phi / \sigma_R$ smaller than $\sqrt{2}/2$.
It may be also accounted for if $R_\sigma/R_\rho>>2$ (Fig. 2).

\begin{figure*}
\psfig{figure=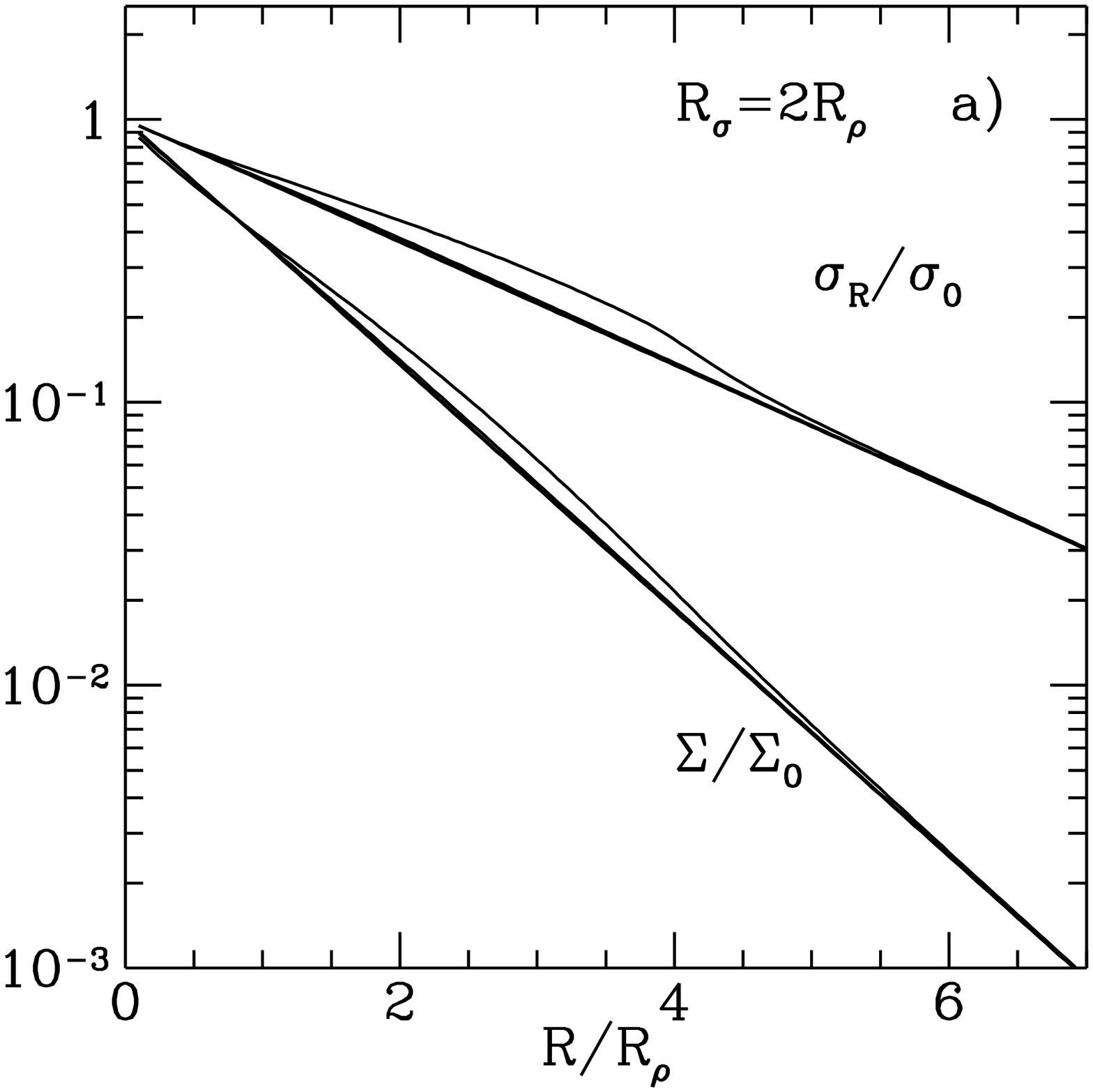,height=5cm,width=5.9cm,angle=0.}
\vspace*{-5.05cm}
\hspace*{5.95cm}
\psfig{figure=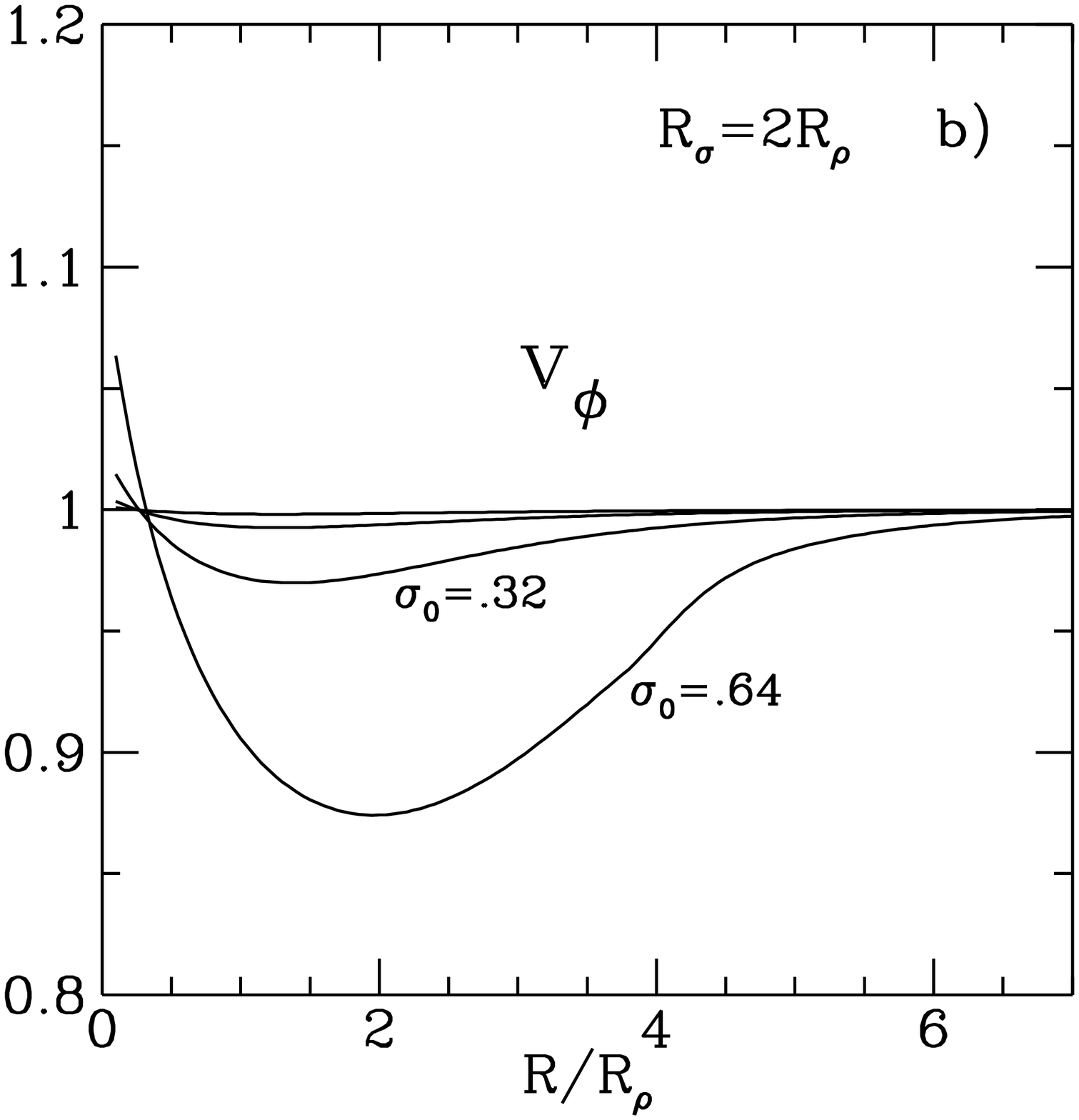,height=5cm,width=5.9cm,angle=0.}
\psfig{figure=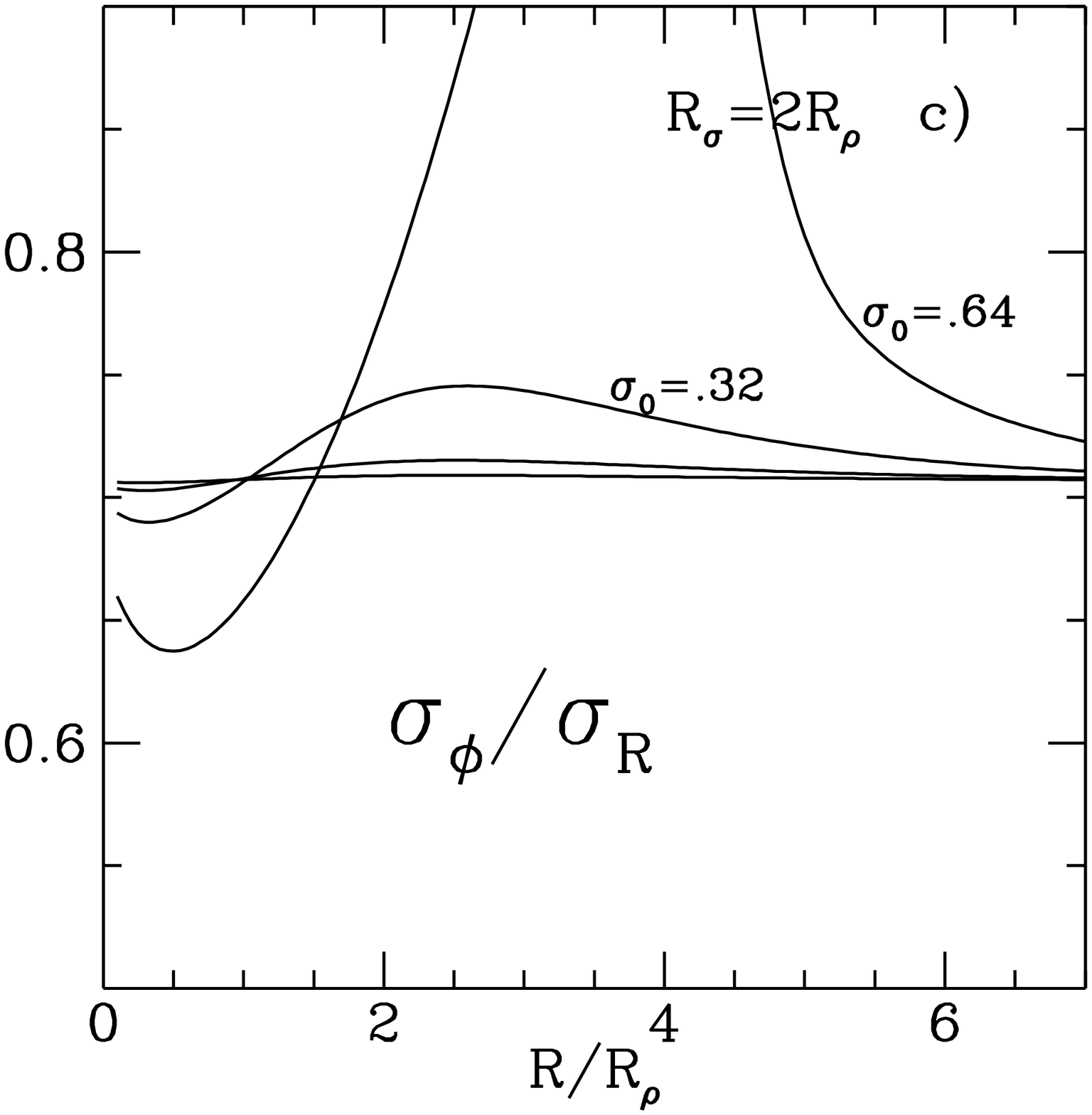,height=5cm,width=5.9cm,angle=0.}
\caption[]{(a) The surface density $\Sigma(R)$ and the radial velocity
dispersion $\sigma_R(R)$ generated in a logarithmic potential by the
distribution of Eqs. 1, 7 and 8 with $R_\sigma=2R_\rho$ and four different
$\sigma_0$ =(.08,.16,.32,.64). (b) The mean streaming velocity. (c) The
velocity dispersion ratio.}
\end{figure*}
\begin{figure*}
\psfig{figure=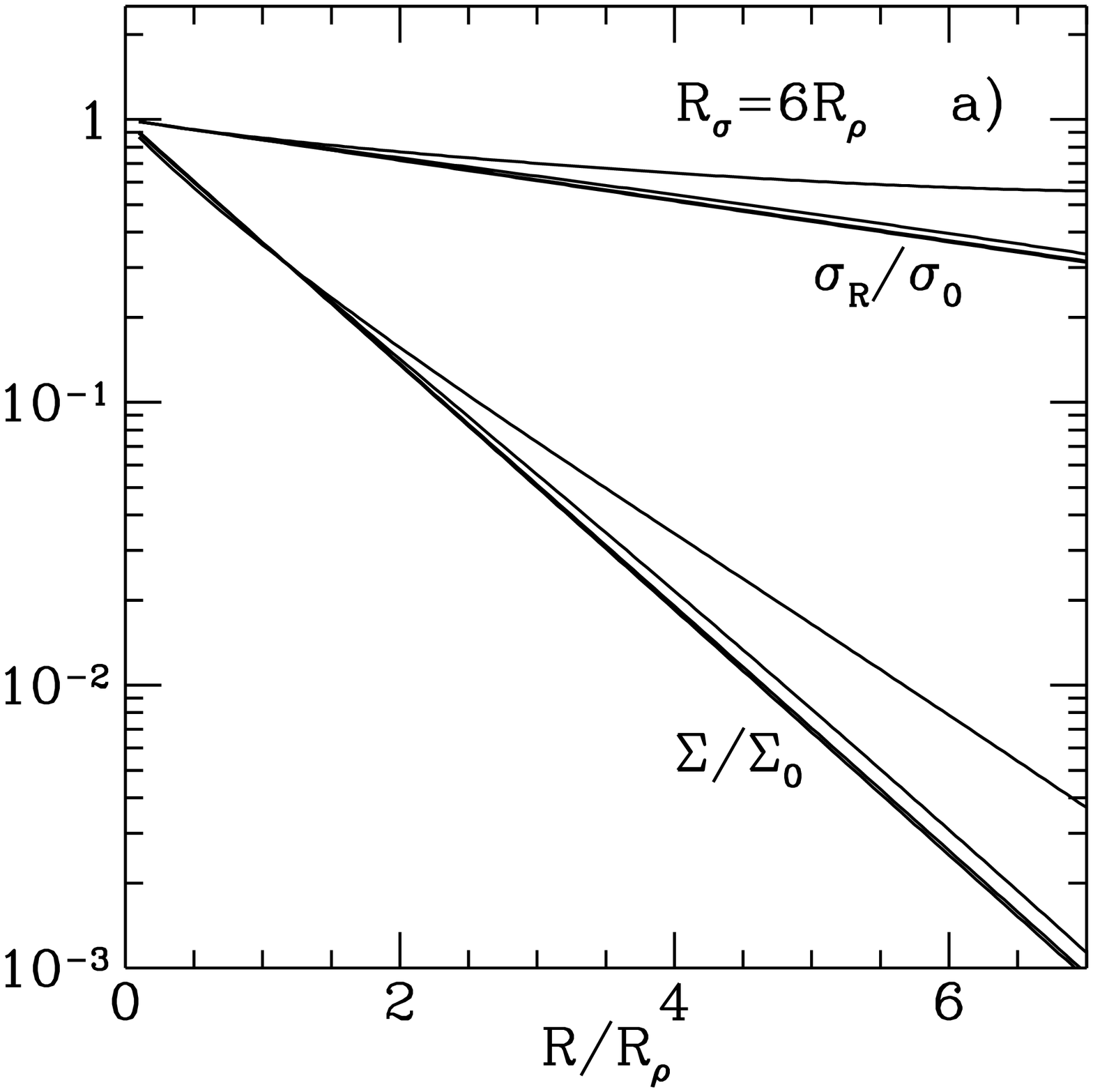,height=5cm,width=5.9cm,angle=0.}
\vspace*{-5.05cm}
\hspace*{5.95cm}
\psfig{figure=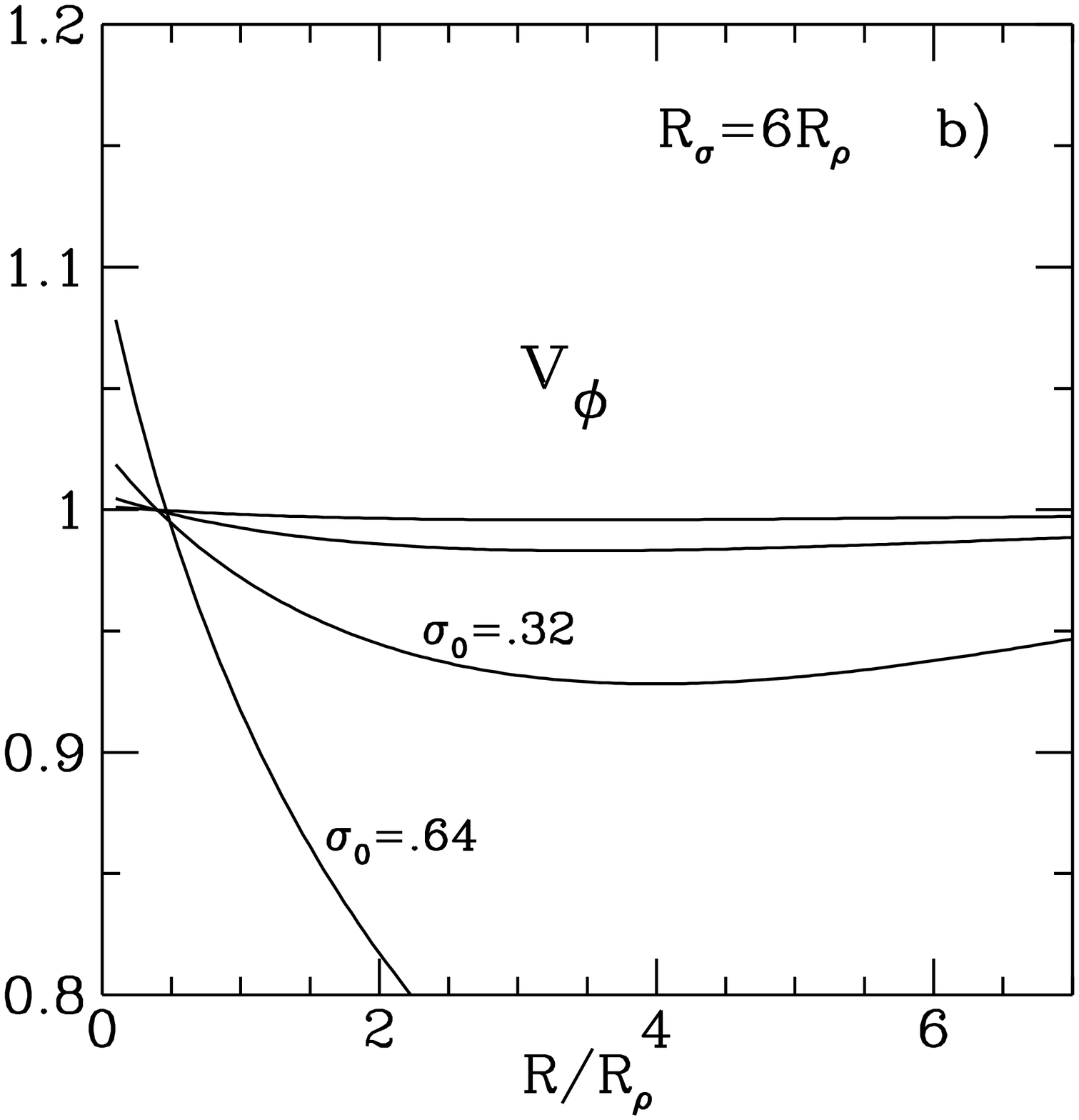,height=5cm,width=5.9cm,angle=0.}
\psfig{figure=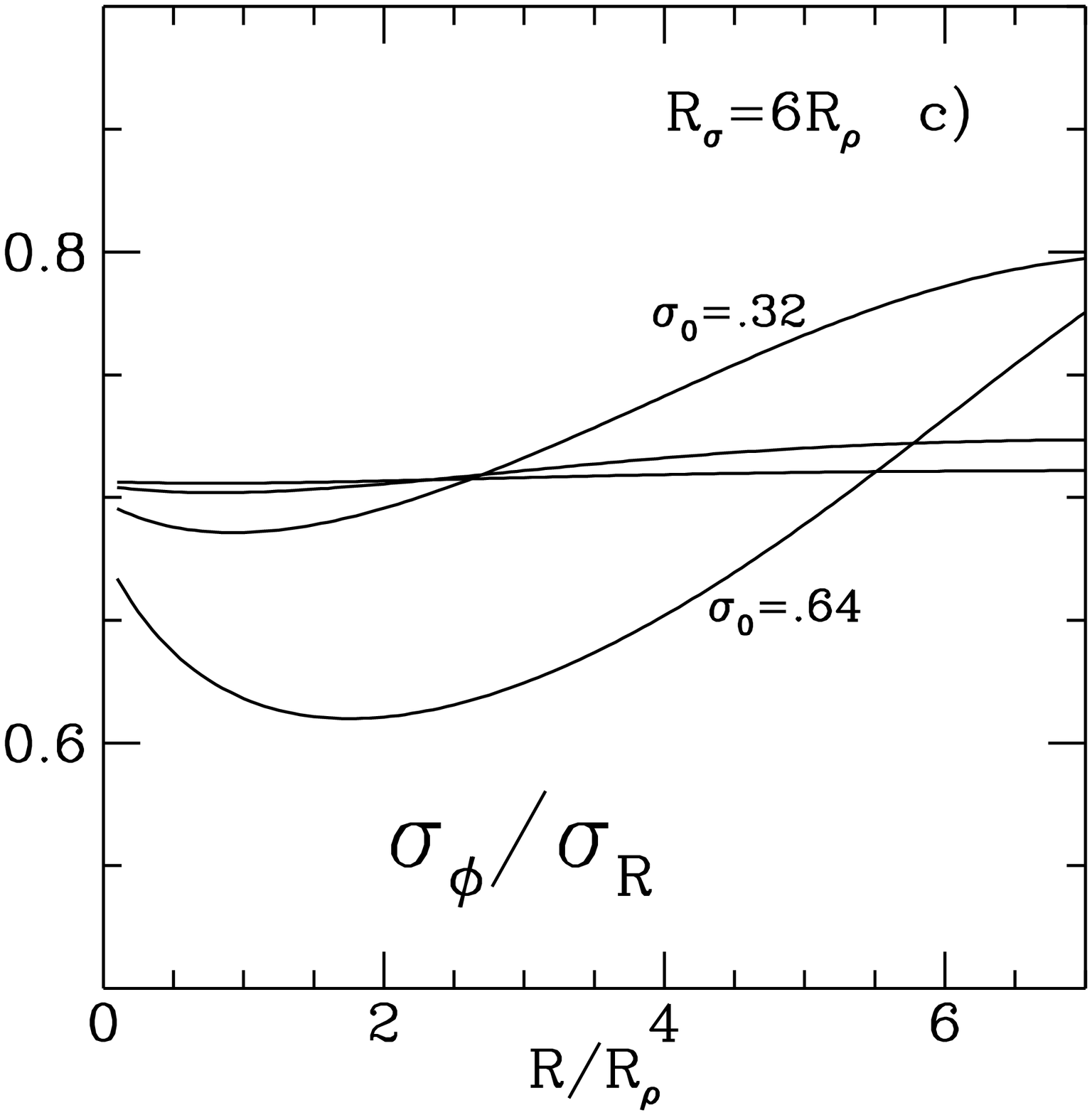,height=5cm,width=5.9cm,angle=0.}
\caption[]{Same as Figure 1, but with $R_\sigma=6R_\rho$.
(a) Surface density and radial velocity dispersion.
(b) The mean streaming velocity. (c) The velocity dispersion ratio.}
\end{figure*}

\subsection{Power law rotation curve models}
In what follows we use a more  general model with potentials yielding a power
law circular velocity curve: $v_c(R)=R^{\alpha}$. The radius of a circular
orbit with momentum $L_z$ is given by
$R_c(L_z)=L_z/v_c=L_z^{ 1 \over \alpha+1}$ and the circular velocity is
$v_c(L_z)=L_z^{\alpha \over \alpha+1}$. This gives for
$\widetilde{\Sigma}$ and $\widetilde{\sigma}$:
\begin{equation}
\widetilde{\Sigma}(L_z) =   \Sigma_0 \exp[-L_z^{1 \over \alpha+1}/R_\rho ]
\end{equation}
\begin{equation}
\widetilde{\sigma}(L_z) = \sigma_0 \exp[-L_z^{1 \over \alpha+1}/R_\sigma ]
\end{equation}

Here again the resulting density and dispersion will be in general nearly
exponential with scale lengths $R_{\rho}$ and $R_{\sigma}$, but models loose
this convenient behaviour when the velocity dispersions are too high.

Other Shu models could be built using different closed forms or using
some spline functions to fit data obtained along the galactic
radius. Such phenomenological models are certainly less elegant than
for example Evans \& Collett (1993) exact models that give DFs for a
few defined density or dispersion distributions. However Shu models
are more flexible, cover a larger range of model types and they have a
sufficient number of free parameters for basic comparison to
observations. Thus they allow a more extensive comparison to
available data.

\subsection{3D models}
3D models are built to include the vertical structure of
the potential. We detail two cases for which explicit DFs are easy to
obtain.  In the case of a plane parallel potential, the DF may be
multiplied by the trivial term:
\begin{eqnarray}
{1\over(2\pi)^{1/2}} {1\over\widetilde{\sigma_z}(L_z) }
exp^{-{\dot{z}^2\over 2 \sigma_z^2}}~~~~.
\end{eqnarray}

For a spherical potential, a 3 dimensional DF may be deduced from the DF given by
Equ. (1) and from the equation 11b given by
Statler (1989) for St\"ackel potentials; in the case of a flat rotation
curve, the vertical term is:
\begin{eqnarray}
{1\over(2\pi)^{1/2}} {1\over\widetilde{\sigma_z}(L_z) }
exp^{
-{\dot{z}^2\over 2}
\left[ {1\over\sigma^2}+({1\over\sigma_z^2}-{1\over\sigma^2})
{ v_c^2 \over v^2 }  \right]}~~~~.
\end{eqnarray}
This shows that the correlation term is the most different for the
highest velocity dispersion populations. In the next section we
consider plane parallel and spherical models with power law rotation
curves. Other more general models using DFs in St\"ackel potentials
(Statler 1989) have also been used to examine solutions in
intermediate potentials between plane-parallel and spherical ones.  In
the present context DFs in St\"ackel potentials introduce a new
parameter: the focus of the ellipsoidal system of coordinates
(Statler, 1989). For all these models we have assumed that the
(radial) vertical and (radial) horizontal dispersion kinematic scale
lengths are equal $R_{\sigma_z}=R_\sigma$.  The results of these
intermediate models are presented in Section 4.3 and can be compared
to the plane-paralel model solutions.

\section{Models versus Observations}
\subsection{Observational data}
In this section we compare the
models presented in the previous section to the local stellar
kinematics as given by two catalogues.
The first one is the recently published Reid et al. (1995) catalogue
that improves considerably the data for K and M stars found in the
CNS3 catalogue of nearby stars (Gliese \& Jahreiss 1991). They
include new radial velocity measurements and reject stars with incorrect
spectral types or giants previously misclassified as dwarfs.

We build a second catalogue by extending as much as possible the CNS3
catalogue using the Simbad database at the CDS in Strasbourg and by
completing missing data, colours, radial velocities and some
trigonometric parallaxes when new observations are available (mainly
from the Hipparcos Input Catalogue, Turon et al. 1992) but also from a
bibliography of
radial velocities (Barbier-Brossat et al. 1994) and the General
Catalogue of Trigonometric Stellar Parallaxes (Van Altena et al.
1991). We remove misclassified stars identified by Reid and stars with
$M_v>8$. We retain only O to G type stars to get a catalogue complementary
to Reid's catalogue.

We extract from these catalogues distance-limited samples in order to
obtain homogeneous and kinematically unbiased 3D velocity distributions. We
use distance criteria given by Reid et al. (1995) for their catalogue, and
apply a similar process to define the completeness limits for the second catalogue
($26~\rm{pc}$ for $M_v<3$, $23~\rm{pc}$ for $3<M_v<5$ and
$21~\rm{pc}$ for $5<M_v<8$).

Finally we consider multiple stars as one kinematic object since the kinematics
of the members are correlated. We reject one star with a velocity modulus larger
than $200\,\kms$ that is certainly a halo star unmodelled by our DF.

Stars or data not in the CNS3 and Reid catalogues are given in Tables 1 and 2.

\begin{table*}
\caption[]{Stars with new measurements labelled with *. Other data are taken
from the CNS3. New V$_{rad}$ are extracted from Barbier-Brossat
et al. (1994), trigonometric parallaxes from
the General Catalogue of Trigonometric Stellar Parallaxes
(Van Altena et al. 1991) and B-V from the Hipparcos Input Catalog
(Turon et al. 1992).}
\begin{flushleft}
\begin{tabular}{lrrcrrlrrrrrr}
\hline
\multicolumn{1}{c}{Identification} &  \multicolumn{1}{c}{$\alpha_{1950}$}
& \multicolumn{1}{c}{$\delta_{1950}$}
& \multicolumn{1}{c}{$15\mu_\alpha\cos\delta$}
& \multicolumn{1}{c}{$\mu_\delta$} & \multicolumn{1}{c}{ V$_{rad}$}
& \multicolumn{1}{c}{Sp} & \multicolumn{1}{c}{$M_v$}
& \multicolumn{1}{c}{B-V} & \multicolumn{1}{c}{$\pi$}
& \multicolumn{1}{r}{u}&\multicolumn{1}{r}{v}&\multicolumn{1}{r}{w} \\
\multicolumn{1}{c}{number} &  \multicolumn{1}{c} {}& \multicolumn{1}{c}{}
& \multicolumn{1}{c}{$\arcsec/y$} & \multicolumn{1}{c}{$\arcsec/y$}
& \multicolumn{1}{c}{km.s$^{-1}$} & & & & \multicolumn{1}{c}{mas}
& \multicolumn{3}{c}{km.s$^{-1}$}\\
\hline
HD~~~8357&01 20 20&07 09.3&~0.095& 0.239&  0.0*&G5  &6.43&     & 66.0~&-13&  8& 11\\
Gl 174.1A&04 38 57&-41 57.5&-0.198&-0.076& -0.6*&F2 V&2.70&0.34~& 44.8~& 11&  7&-11\\
 HD 30090&04 42 56&42 15.5&~0.000& 0.065& 28.8~&G0  &5.10&0.70*& 52.0~&-26& 13&  3\\
 HD 38114&05 41 35&32 22.3&~0.020&-0.083&-44.4*&G5  &6.56&     & 48.0~& 44&-10& -4\\
Gl 244~~A&06 42 57&-16 38.8&-0.570&-1.210& -7.6*&A1 V&1.47&0.00~&380.4~& 14&  0&-11\\
Gl 253   &06 53 51&-55 11.5&-0.042&-0.180& 40.1~&G7 V&7.38&0.78~& 69.6*&  8&-36&-20\\
Gl 274~~A&07 25 54&31 53.1&~0.181& 0.173& -4.4*&F0 V&3.04&0.32~& 59.1~&  9& 10& 14\\
Gl 291~~A&07 49 27&-13 45.8&-0.069&-0.344&-17.8*&F9 V&4.71&0.57~& 62.7~& 25&  0&-19\\
Gl 292.2 &07 52 03&-01 16.8&-0.278&-0.054& 93.4*&G8 V&5.60&0.73~& 51.9~&-77&-58& -2\\
HD 108081&12 22 25&-03 56.7&-0.144&-0.214& 47.0~&G5  &6.80&0.66*& 54.7~&  7&-43& 29\\
HD 163621&17 53 40&36 11.7&-0.145&-0.015&  0.0*&G5  &6.33&     & 50.0~&  2& -6&  9\\
\hline
\end{tabular}
\end{flushleft}
\end{table*}

\begin{table*}
\caption{Supplementary stars. Data obtained or deduced from the Hipparcos
Input Catalog (Turon et al. 1992).}
\begin{tabular}{rrrcrrlrrrrrr}
\hline
\multicolumn{1}{c}{HIC} &  \multicolumn{1}{c}{$\alpha_{1950}$}
& \multicolumn{1}{c}{$\delta_{1950}$}
& \multicolumn{1}{c}{$15\mu_\alpha\cos\delta$}
& \multicolumn{1}{c}{$\mu_\delta$}
& \multicolumn{1}{c}{ V$_{rad}$} & \multicolumn{1}{c}{Sp}
& \multicolumn{1}{c}{$M_v$} & \multicolumn{1}{c}{B-V}
& \multicolumn{1}{c}{$\pi$} & \multicolumn{1}{r}{u}&\multicolumn{1}{r}{v}
& \multicolumn{1}{r}{w} \\
\multicolumn{1}{c}{number} &  \multicolumn{1}{c} {}& \multicolumn{1}{c}{}
& \multicolumn{1}{c}{$\arcsec/y$} & \multicolumn{1}{c}{$\arcsec/y$}
& \multicolumn{1}{c}{km.s$^{-1}$} & & &
& \multicolumn{1}{c}{mas}& \multicolumn{3}{c}{km.s$^{-1}$}\\
\hline
  3405&00 41 07&-57 44.2 &-0.012&  0.011 &   2.0 & A0 IV& 2.36 & 0.02 &39.0&     1&    1&   -2  \\
  4422&00 53 40&58 54.7 &-0.174& -0.048 & -47.6 & G8 IV& 2.64 & 0.96 &40.0&    35 & -34&   -3  \\
  6607&01 22 27&-28 05.7 &~  0.359& -0.287 &  87.2 & G6 V & 7.46 & 0.68 &68.0&   -14&  -36&  -83 \\
  9381&01 58 25& -40 58.0 &~  0.593& -0.436 & -27.5 & G3 V & 5.99 & 0.65 &   53.0&    -1&  -43&   45 \\
 20347&04 19 27& -25 50.7 &~ 0.056& -0.056 &  18.0 & F2 V & 4.48 & 0.35 &   49.0&    -7&  -15&  -10 \\
 26834&05 39 38& -15 39.1 &~  0.232& -0.106 &  67.0 & G8  & 6.74 & 0.76 &   75.0&   -43&  -51&  -16  \\
 49669&10 05 43& 12 12.7 & -0.254&  0.006 &   5.9 & B7 V &-0.68 &-0.09 &   39.0&   -27&   -7&  -13 \\
 55642&11 21 19& 10 48.3 &~ 0.170& -0.075 & -10.3 & F2 IV& 2.36 & 0.42 &   47.0&    20&    4&   -6 \\
 57606&11 46 04& 14 33.7 & -0.110&  0.000 &   3.0 & F0 V & 4.26 & 0.30 &   47.0&   -10&   -5&    0 \\
 76267&15 32 34& 26 52.9 &~0.136& -0.089 &   1.7 & A0 V & 0.39 &-0.01 &   43.0&    15&    3&   -7 \\
 81833&16 41 11& 39 01.0 &~0.048& -0.082 &   8.3 & G8 IV& 2.11 & 0.92 &   53.0&    10&    5&    2 \\
 93519&19 00 11& -00 47.1 &~0.031& -0.011 & -26.0 & G5  & 7.38 & 0.72 &   52.0&   -22&  -14&   -2  \\
 98066&19 52 47& -26 26.0 &~0.231&  0.083 & -18.6 & G3  & 3.40 & 0.75 &   55.0&   -25&    6&   -5  \\
 98258&19 55 07& -15 37.5 &~0.017& -0.100 &   3.0 & A2 V & 3.32 & 0.05 &   46.0&     4&   -8&   -6 \\
105864&21 23 55& 00 53.3 &~0.113& -0.163 &  11.0 & F5 V & 4.63 & 0.45 &   50.0&     5&   -4&  -21 \\
113136&22 51 60& -16 05.2 & -0.041& -0.025 &  18.0 & A3 V & 1.24 & 0.07 &   39.0&    11&    5&  -14 \\
\hline
\end{tabular}
\end{table*}

\subsection{Model adjustment}
Distribution functions are modeled by Eqs. 1, 2, 7 and 8. Data-model
comparison is done by adjustment of model parameters using the maximum
likelihood method in the ($v_R$, $v_\phi$) space for plane-parallel
models (and $v_R$, $v_\phi$, $v_z$ for 3D models). All stars are
assumed to be at solar galactic radius $R=R_0$. The
$u_{\odot},~v_{\odot}$ components of the solar velocity relative to
the LSR are two local parameters. Other parameters $\Sigma_0$,
$\sigma_0$ are dependent upon the local number of stars and the local
radial velocity dispersions.  Non-local parameters are the kinematic
and the density scale lengths of stellar populations and the slope of
the circular velocity curve. These quantities are obtained as a mean
over a few kiloparsecs around the solar galactic radius corresponding
to the extent of the radial excursions of stars that are in the solar
neighbourhood, and that collectively carry some information from these
more distant regions.  For 3D models, the coordinate of
the focus of the ellipsoidal coordinate system used to define the
St\"ackel potential is a supplementary adjustable parameter.

The radial to tangential velocity dispersion ratio of a low dispersion
population constrains mainly the slope of the circular velocity
curve. For kinematically hot populations, this ratio depends also on
$R_\rho$ and $R_\sigma$. The mean velocity $\overline{v}_\phi$
is determined by asymmetric drift. Comparing stellar groups with
small and large velocity dispersions gives a more accurate measurement
of asymmetric drift.  For this reason we divide the catalogue in
samples (identified by a number k) according to the absolute
magnitudes. The sample including the smallest absolute magnitude
corresponds to the brightest stars or the bluest since all these stars are
on the main sequence. This sample includes young stars and
has their kinematic characteristic: low velocity dispersion.
In contrast, samples with fainter stars have redder
main sequence stars and are a mixture which includes older stars with higher
velocity dispersion. We determine the sample division by minimizing the
errors on the $R_\rho$ and $R_\sigma$ estimates.
%
\begin{figure}
\psfig{figure=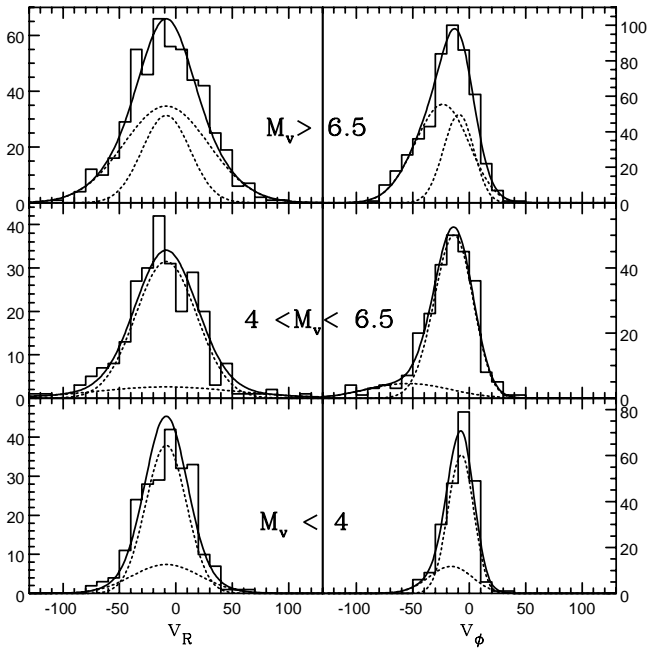,width=8.8cm,bbllx=45pt,bblly=160pt,bburx=590pt,bbury=700pt,clip=true}
\caption[]{
Radial and tangential velocity histograms for three samples of stars sorted
according to their absolute magnitude $M_v$. The maximum likelihood models are
plotted (continuous lines) as well as individual DFs (dashed lines).
}
\end{figure}
\\
Maximum likelihood is used to estimate the best model parameters.
The likelihood function $L$ is the product of likelihoods $L_k$, where
$L_k$ is the likelihood of sample number k.
\begin{equation}
L_k = \prod_{i=1,N_k} f_k(v_{R,i},v_{\phi,i}) /\Sigma_k
\end{equation}
with
\begin{equation}
\Sigma_k= \int\int f_k dv_{R} dv_{\phi}
\end{equation}
where $N_k$ is the number of stars in sample k.

Each sample is modeled by using two elementary distribution functions,
$f_k=f_{k,1}+f_{k,2}$ (continuous lines in Fig. 3) corresponding to the sum
of two DFs (dashed lines in Fig. 3) given by Eqs. (1, 2, 7 and 8).

\subsection{Results}
Solutions do not depend significantly on the catalogues, the splitting
method or the shape of the potentials considered (spherical,
plane-parallel or intermediate). Best fits with the smallest errors
are obtained using both catalogues simultaneously. Figure 3 shows the
observed velocity histograms and the corresponding model curves for
the best estimates of parameters with a plane-parallel potential and
$R_0=8.5\,\kpc$ and $V_0=220\,\kms$.  This model is in good agreement
with data (the reduced $\chi^2$ is approximately one for each
histogram).\\ The results obtained with each catalogue are given in
Table 3 in case of plane-parallel models and in case of intermediate
3D models. $R_{\rho,eff}$ and $R_{\sigma,eff}$ scale lengths are the
effective scale lengths of the computed density or dispersion that, in
the case of high velocity dispersions, differ from the $R_\rho$ or
$R_\sigma$ parameters used in Eqs. 3 and 4. Formal errors are small
but we find a strong correlation of scale length with solar velocity
$v_\odot$ which limits the accuracy of density scale length estimates
and could bias our conclusions.

\begin{table*}
\caption{Model parameters: solar velocities, density and kinematic scale
lengths, slope of the galactic velocity curve. Galactic structural parameters
are obtained for various splittings of the data in absolute magnitude and
assuming $R_0=8.5\,\kpc$ and $V_0=220\,\kms$.}
\begin{tabular}{lclccccc}
\hline
Catalogue       & number & splitting & $u_\odot$
& $v_\odot$ & $R_{\rho,eff}$ & $R_{\sigma,eff}$ & $\alpha$ \\
& of stars && $\kms$ & $\kms$ & $\kpc$ & $\kpc$ &\\
\hline
Plane-parallel models\\
\hline
Reid et al.& 477 & 1
&$8.4\pm1.5$ & $4.3\pm4.4$ & $1.9$ & $36$ & $-.21\pm.07$ \\

Blue nearby stars & 504 & 2 (228/276)
&$8.7\pm1.1$ & $2.5\pm2.1$ & $1.8$ & $24-30$ & $-.23\pm.06$ \\

Reid and Blue stars & 981 & 3 (228/256/497)
&$8.6\pm0.9$ & $2.2\pm1.7$ & $1.7\pm.3$ & $>30$ & $-.22\pm.05$ \\

Reid and Blue stars with & 981 & 3 (228/256/497)
& $8.6\pm0.9$ & $2.6\pm1.6$ & $2.0\pm.3$ & $>30$ & $-.18\pm.04$ \\
$R_0=7.9\,\kpc$, $V_0=185\,\kms$ & & & & & & & \\

Reid and Blue stars : {\boldmath $v_\odot=8$} & 981 & 3 (228/256/497)
& $8.7\pm.9$ & {\bf 8} & $2.9\pm.3$ & $18-30$ & $-.20\pm.05$ \\
\hline
3D models\\
\hline

Reid and Blue stars & 981 & 3 (228/256/497)
& $8.5\pm1.2$ & $2.5\pm2.2$ & $1.7\pm.3$ & $>30$ & $-.22\pm.06$ \\

Reid and Blue stars : {\boldmath $v_\odot=8$} & 981 & 3 (228/256/497)
& $8.6\pm1.2$ & {\bf 8} & $2.8\pm.4$ & $20-30$ & $-.20\pm.06$ \\
\hline
\end{tabular}
\end{table*}

\subsection{Sample splitting}
We divided the stellar catalogue into one, two or three samples with various
sizes and we made systematic explorations. We found subdivisions that minimize
errors of non-local parameters.
Within the error bars, other subdivisions gave solutions
in agreement with the best fit.
The best fit is obtained with two small samples with a low velocity dispersion
and a large one with a high velocity dispersion. This corresponds certainly to
the best way to constrain the asymmetric drift.

\subsection{$R_\rho$ and the solar velocity $v_\odot$}
The maximum likelihood solution gives a small value of $R_{\rho,eff}=
1.7\,\kpc$.  Plotting the asymmetric drift relation $v'_\odot=
v_\odot+c<\Pi^2>$ with the CNS3 catalogue yields a result different
from Delhaye (1965) who obtained $v_\odot=12\,\kms$.  It
appears that the mean velocity $v$ of the few tens of bluest CNS3 main
sequence stars is small and favours a small $v_\odot$ value between 3
and $7\,\kms$. This explains the small $R_\rho$ found which is strongly
correlated with $v_\odot$.  Most recent $v_\odot$ determinations
are small, G\'omez and Mennessier (1977) found $6\,\kms$, Mayor (1974)
obtained $6.3\,\kms$ and Delhaye (1982) found $8.5\,\kms$. Some other
references may be found in Kuijken and Tremaine (1991).  Oblak (1983)
built new samples carefully determining stellar ages in order to
analyse the asymmetric drift. He found a small solar velocity
$v_\odot=5.0\pm0.7\,\kms$.

To summarize, if we admit $2$ and $8\,\kms$ as the lower and upper
limits on the $v_\odot$ solar velocity, we obtain limits on the
effective density scale lengths $R_{\rho,eff}$ from $1.7\,\kpc$ to
$2.9\,\kpc$.

The scale length we have determined is proportional to the assumed
solar galactic radius $R_0$; with our best fit it gives $R_\rho=
1.7*(R_0/8.5)\,\kpc$.

\section{Discussion and conclusion}
Dynamical analysis of the stellar kinematics in our Galaxy is
frequently based on the comparison between observed and modelled
moments of the velocity and not directly on the measured distribution
function. A practical reason for that is the apparent lack of simple
dynamical models predicting realistic distribution functions. Moments
contain most of the dynamical information. However their measurements
based on observational data may be strongly biased by a few stars
(high velocity members in a binary or halo stars when analyzing the
disc). For example Cuddeford \& Binney (1992) show how these biases
can be reduced. Evans \& Collett (1993) obtain new realistic DFs by
generalising the Rybicki disc model, but in the case
of exponential disk models with  $R_{\rho}=R_{\sigma^2}$,
they only obtain  moments.

Here we have shown that {\it the Shu (1969) models that are exact solutions
of the collisionless Boltzmann equation, can be used successfully to
analyse galactic dynamics}. We build a nearly exponential stellar disc
(much more accurately exponential than galactic discs really are),
which includes the basic necessary parameters, the density and
kinematic scale lengths, the slope of the rotation curve and which
removes all restraining assumptions done in previous work.  The
comparison to the velocity distribution from an unbiased kinematic
sample of nearby stars gives an accurate determination of the galactic
structure: assuming $R_0=8.5\,\kpc$ and $V_0=220\,\kms$, {\it we conclude
that the stellar density scale length of the Milky Way ranges between
1.7 and 2.9 $\kpc$}. We obtain a large kinematic scale length; this is
in poor agreement with the galactic $R_\sigma$ determination
obtained by Lewis and Freeman (1989) or by Bottema (1993) in external
similar spiral galaxies. The large kinematic scale length that we find
may result from the large proportion of young stars in our sample with
a kinematic behaviour resembling gas that has a null kinematic
gradient. It may result from the inadequacy of the model
to also measure this parameter, e.g. if the accuracy
of velocities in the Gliese catalogue is not sufficient. This could be
checked swiftly with Hipparcos data. We find that {\it the rotation curve
$v_c(R)=R^\alpha$ is decreasing at the solar galactic radius with
$\alpha=-0.22\pm.05$}. All these quantities are obtained from local
stars that have non-negligible radial motions. It means that these
structural parameters are mean values measured over an extent of a few
kiloparsecs around the Sun.  Mayor \& Oblak (1985) and Oblak \& Mayor
(1987) proceed partly with a similar approach (maximum likelihood and
Monte-Carlo) but based on numerical integration of stellar orbits in
different galactic potentials. They find $R_{\sigma^2}=5\,\kpc$ or
$R_{\sigma}=10\,\kpc$.

Few other previous kinematic determinations of the Galaxy scale lengths
have been obtained assuming $R_{\rho}=R_{\sigma^2}$. Determinations were then
obtained directly from the measurement of the kinematical scale length in the
mid-plane, or from the asymmetric drift relation. Here the analysis of the
local kinematics of stars shows that this assumption is wrong in the Galaxy
and introduces bias favouring much larger estimates of $R_\rho$.

Recent analyses, using simpler distribution functions (Evans \& Collett 1993
and Cuddeford \& Binney 1992), one assuming $R_{\rho}=R_{\sigma^2}$ the other
a flat rotation curve, were not able to reproduce accurately the observed DF.
Our results are concordant with the work by Fux \& Martinet (1994) based on
Jeans equations expanded by Cuddeford \& Amendt (1991) with assumptions to
obtain the closure of the moment equations. They deduce a short density scale length
and they conclude, as we do here, that this result does not support the
assumption $\sigma_z(R)/\sigma_R(R)=const.$

The structural parameter values are obtained with a good accuracy. The
small value of $R_\rho$ obtained with our analysis of the kinematics of
neighbouring stars is consistent with most recent determinations of the density:
Robin et al. (1992ab) and Ojha et al.
(1996) based on star counts or Durand et al. (1996) based on COBE maps. {\it We have
shown that the hypothesis relating density and kinematic scale length
$R_\rho=R_\sigma/2$ is incorrect} and that scale length determinations based on this
assumption must be ruled out.

\acknowledgements{We would like to thank C. Lineweaver for his careful
reading of the manuscript and L. Martinet for his valuable help to
improve the paper.  \\This research has made use of the Simbad database,
operated at CDS, Strasbourg, France.}


\begin{thebibliography}{}


\bibitem{b1} Barbier-Brossat M., Petit M., Figon P., 1994, A\&AS, 108, 603

\bibitem{b2} Blitz L., Spergel D.N., 1991, ApJ, 370, 205

\bibitem{b3} Bottema R., 1993, A\&A, 275, 16

\bibitem{b4} Burton W.B., te Lintel Hekkert P., 1986, A\&AS, 65, 427

\bibitem{c1} Cuddeford P., Amendt P., 1991, MNRAS, 253, 427

\bibitem{c2} Cuddeford P., Binney J., 1994, MNRAS, 266, 273

\bibitem{d1} Dambis A.K., Mel'nik A.M., Rastorguev A.S., 1995,
Astronomy Letters Vol. 21 No 3, 291

\bibitem{d2} Dehnen W., Binney J., 1995, ''Formation of the Galactic Halo''
Tucson, October 9-11, 1995 ASP Conference Series, eds Sarajedini A.
\& Morrison H.

\bibitem{d3} de Jong R.S., 1996, A\&A, 313, 377

\bibitem{d4} Delhaye J., 1965, in  Galactic Structure, Blaauw A., Schmidt M.,
eds, Univ. Chicago Press, Chicago, ch.4

\bibitem{d5} Delhaye J., 1982, Mitteilungen der Astron. Gesellschaft, 57, 123

\bibitem{d6} Durand S., Dejonghe H., Acker A., 1996, A\&A, 310, 97

\bibitem{e1} Elmegreen D.M., Elmegreen B.G., 1984, ApJS, 54, 127

\bibitem{e2} Evans N.W., Collett J.L., 1993, MNRAS, 264, 353

\bibitem{f1} Fich M., Blitz L., Stark A.A., 1989, ApJ, 342, 272

\bibitem{f2} Fux R., Martinet L., 1994, A\&A, 287, L21

\bibitem{g1} Giovanardi C., Hunt L.K., 1988, AJ, 95, 408

\bibitem{g2} Gliese W., Jahreiss H., 1991, Preliminary Version of the Third
Catalogue of Nearby Stars, on The Astronomical Data Center CD-ROM: Selected
Astronomical Catalogs, Vol. 1, eds. L. E. Brotzmann and S.E. Gesser,
NASA/ADC, Greenbelt, MD

\bibitem{g3} G\'omez, A., Mennessier M.O., 1977, A\&A, 54, 113

\bibitem{h1} Haywood M., Robin A.C., Cr\'ez\'e M., 1996, A\&A (in press)

\bibitem{k1} Kent S.M., Dame T.M., Fazio G., 1991, ApJ 378, 131

\bibitem{k2} Kerr F.J., Lynden-Bell D., 1986, MNRAS, 221, 1023

\bibitem{k3} Kuijken K., Gilmore G., 1989, MNRAS, 239, 605

\bibitem{k4} Kuijken K., Tremaine S., 1991, {\sl Dynamics of Disk Galaxies},
Ed: B. Sundelius (G\"oteborg Univ. Press), 71

\bibitem{k5} Kuijken K., Tremaine S., 1994,  ApJ, 421, 178

\bibitem{l} Lewis J.R., Freeman K.C., 1989, AJ, 97, 139

\bibitem{m1} Mayor M., 1974, A\&A, 32, 321

\bibitem{m2} Mayor M., Oblak E., 1985, The Milky Way Galaxy, Symp IAU 106,
eds: H. van Woerden et al., 149

\bibitem{n} Neese C.L., Yoss K.M., 1988, AJ, 95, 463

\bibitem{o1} Oblak E., 1983, A\&A 123, 238

\bibitem{o2} Oblak E., Mayor M., 1987, Evolution of Galaxies, X IAU
European meeting, ed. J. Palous, Publ. Astron. Inst. Czech. Acad. Sci. 69, 263

\bibitem{o3} Ojha D.K., Bienaym\'e O., Robin A.C., Cr\'ez\'e M., Mohan V.,
1996, A\&A 311, 456

\bibitem{p} Pont F., Mayor M., Burki G., 1994, A\&A, 285, 415

\bibitem{r1} Reid I.N., Hawley S. L., Gizis J.E., 1995, AJ, 110(4), 1838

\bibitem{r2} Reid M.J., 1993, ARA\&A, 31, 345

\bibitem{r3} Robin A.C., Cr\'ez\'e M., Mohan V., 1992a, A\&A, 265, 32

\bibitem{r4} Robin A.C., Cr\'ez\'e M., Mohan V., 1992b, ApJ, 400, L25

\bibitem{r5} Rohlfs K.,  Kreitschmann J., 1988, A\&A, 201, 51

\bibitem{s1} Shu F.H., 1969, ApJ, 158, 505

\bibitem{s2} Statler T.S., 1989, ApJ 344, 217

\bibitem{t1} Tamisier F., 1991, {\sl Rapport de stage de DEA},
Observatoire de Paris

\bibitem{t2} Turon C., Cr\'ez\'e M., Egret D. et al., 1992, The Hipparcos
Input catalogue, Ed: ESA Publication division c/o ESTEC, Noordwijk

\bibitem{v1} van Altena W.F., Lee J.T., Hoffleit D., 1991,
The General Catalogue of Trigonometric Stellar Parallaxes, Preliminary
Version (catalogue available at CDS: number I/174)

\bibitem{v2} van der Kruit P.C., 1986, A\&A, 157, 230

\end{thebibliography}
\end{document}